\documentclass[11pt,conference]{IEEEtran}
\usepackage{cite}

\usepackage{textcomp}
\usepackage{hyperref}
\usepackage{footnote}
\usepackage{booktabs}
\usepackage{multirow}
\ifCLASSINFOpdf
  \usepackage[pdftex]{graphicx}
  \DeclareGraphicsExtensions{.pdf,.jpg,.png}
\else
\fi

\newcommand{\specialcell}[2][c]{%
  \begin{tabular}[#1]{@{}l@{}}#2\end{tabular}}

\usepackage[tight,footnotesize]{subfigure}

 \renewcommand{\baselinestretch}{0.978}

\usepackage[justification=centering]{caption}
\begin{document}
%
\title{TLB and Pagewalk Performance in Multicore Architectures with Large Die-Stacked DRAM Cache}

\author{\IEEEauthorblockN{Adarsh Patil}
\IEEEauthorblockA{ \textit{Computer Science and Automation}\\
\textit{Indian Institute of Sciences, Bangalore, IN}\\}
\textit{adarsh.patil@csa.iisc.ernet.in} 
}


%


\maketitle

\begin{abstract}
In this work we study the overheads of virtual-to-physical address translation in processor architectures, like x86-64, that implement paged virtual memory using a radix tree which are walked in hardware.\\
Translation Lookaside Buffers are critical to system performance, particularly as applications demand larger memory footprints and with the adoption of virtualization; however the cost of a TLB miss potentially results in multiple memory accesses to retrieve the translation. Architectural support for superpages has been introduced to increase TLB hits but are limited by the operating systems ability to find contiguous memory. Numerous prior studies have proposed TLB designs to lower miss rates and reduce page walk overhead; however, these studies have modeled the behavior analytically. Further, to eschew the paging overhead for big-memory workloads and virtualization, Direct Segment maps part of a process\textquotesingle \ linear virtual address space with segment registers albeit requiring a few application and operating system modifications.\\
The recently evolved die-stacked DRAM technology promises a high bandwidth and large last-level cache, in the order of Gigabytes, closer to the processors. With such large caches the amount of data that can be accessed without causing a TLB fault - the reach of a TLB, is inadequate. TLBs are on the critical path for data accesses and incurring an expensive page walk can hinder system performance, especially when the data being accessed is a cache hit in the LLC.\\
Hence, we are interested in exploring novel address translation mechanisms, commensurate to the size and latency of stacked DRAM. By accurately simulating the multitude of multi-level address translation structures using the QEMU based MARSSx86 full system simulator, we perform detailed study of TLBs in conjunction with the large LLCs using multi-programmed and multi-threaded workloads.\\
\end{abstract}


%
\IEEEpeerreviewmaketitle

\section{Introduction}
Memory Management Units (MMU) have historically divided the virtual address space into pages, each usually being a few kilobytes. The lower bits are the offset within a page and the upper bits form the virtual page numbers which are translated into corresponding physical page number in main memory. Address Translations use in-memory tables called page tables to map virtual page numbers to physical page numbers. The page table entries also contain meta-information regarding the page including dirty bit, bits for replacement, access privilege, cacheable page etc. A cache of these entries is stored in a special CAM structure called the Translation Lookaside Buffers (TLB). \par

In the early x86 processors, from the Intel 80386 to the Pentium, the page table had at most two levels, which meant that on a TLB miss at most two memory accesses were needed to complete the translation. However, as the physical and virtual address spaces supported by x86 processors have grown in size, the maximum depth of the tree has increased, to three levels in the Pentium Pro to accommodate a 36-bit physical address within a page table entry, and recently to four levels in the AMD Opteron to support a 48-bit virtual address space \cite{amd48bit}. TLBs are accessed on every instruction and data reference and a TLB miss overhead can adversely impact system performance, incurring multiple memory accesses to complete the translation. Furthermore, modern application’s are heavily data centric and have larger memory footprint, as evidenced in big data \cite{bigdatabench} and scale-out cloud applications\cite{cloudsuite}, there is increased pressure on the TLBs to have a larger "TLB-reach" to remain effective. Under virtualization environment, where the guest’s view of physical memory is different from system’s view of physical memory, a second level of address translation is required to convert guest physical addresses to machine addresses which further increases the overhead of a TLB miss \cite{amd2d}. \par
\begin{figure}[h!]
\centering
\includegraphics[scale=0.5]{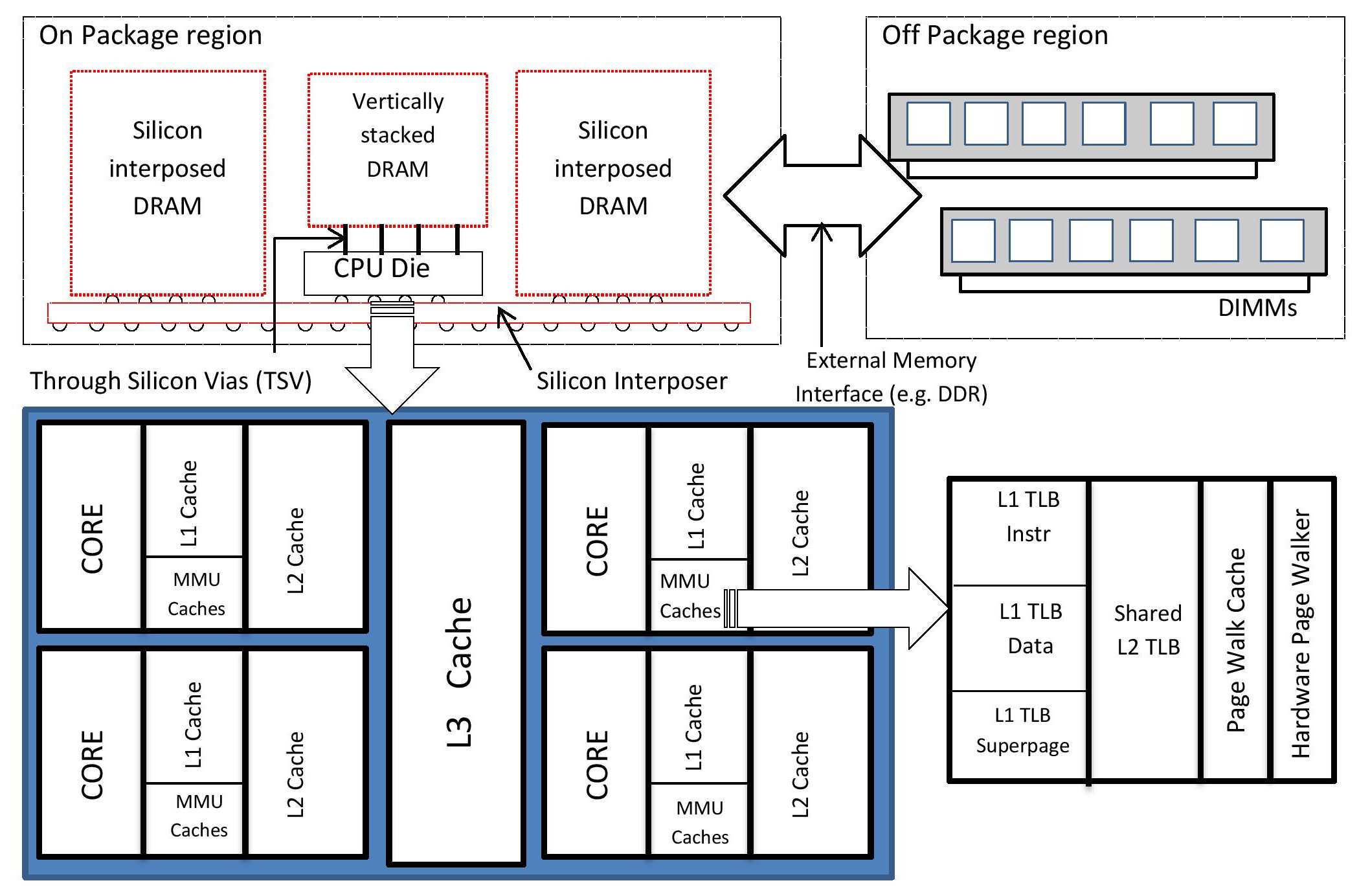}
\caption{Modern processor organization showing MMU caching structures and proposed Die-stacking technology}
\label{arch}
\vspace{-1.5em}
\end{figure}
Parallely, DRAM memory speeds have not been able to keep pace commensurate to CPU  speeds which has led to the well known "Memory Wall" bottleneck\cite{memorywall}. This has led to development of complex memory hierarchies with replication, to avoid off-chip accesses. Also, currently the DRAM memory chips are fabricated using the high-density NMOS process to create high-quality capacitors and low-leakage transistors while logic chips are manufactured using high-speed CMOS process to create complex multi-level metalizations for transistors. The two processes are not compatible to be interfaced on the same die and must be interfaced using off-chip interconnects that add to the latency of an access. \par

The advent of die-stacking technology \cite{diestacking} provides a way to integrate disparate silicon die with better interconnects. The implementation could be accomplished by 3D vertical stacking of DRAM chips using through-silicon vias (TSV) interconnects or horizontally/2.5D stacking on a interposer chip as depicted in Figure \ref{arch}. This allows the addition of a sizable DRAM chip close to processing cores. The onchip DRAM memory can provide anywhere from a couple of hundreds of megabytes to a few gigabytes of storage at high bandwidths of ~400GB/s compared to the 90GB/s of DDR4 bandwidth. This on-chip memory has been advocated to be used as a large last level cache which is transparent to software. In this context, incurring expensive page walks on TLB miss, when the data being accessed is a hit in these large LLCs can hinder system performance. \par

This work studies and explores the following design opportunities:
\begin{itemize}
\item We study the problem of reach of the multitude of TLBs in the modern processors and the associated latency of page walks on a TLB miss in detail. We find that some workloads suffer from significant page walk overheads, incurring as much as 135 cycles on an average per dTLB miss. This overhead is high compared to 40-50 cycles for accessing die-stacked LLC or 80-100 cycles for off-chip DRAM memory access.
\item We quantify the effects of caching of page walk levels in the cache hierarchy and the TLB miss overheads on a modern OoO processor architecture. We find that for the higher levels of the page walk tree, only about 20\%-30\% of accesses hit in L1 or L2 cache. The lower levels are found either in L3 caches or access main memory for translation. \item We also evaluate the impact on IPC with ideal TLB / address translation in an OoO processor. The total overheads of address translation system on IPC values for some workloads is as much as 10\% 
\item Finally, we correlate the TLB reach to the size and latency of the large last level die-stacked caches. Across a wide range of cache sizes and 2 different block sizes, we analyze the efficiency of TLBs to use the proposed large capacity caches efficiently.
\end{itemize}

Overall, this work is an early study towards the design of efficient MMUs and caches for future multi-core designs. In particular our results are focused on understanding the TLB-reach for the large die-stacked DRAM cache to avoid needless page walk overheads when data accessed is present in caches close to the processor.\par 
The rest of the paper is organized as follows. Section II introduces the relevant aspects of page table structure in modern x86-64 architecture and describes the page table walk in native and virtualized environments. Section II also describes the die-stacked DRAM caches. Section III discusses related work. Section IV lists our simulator infrastructure and experimental methodology. Section V presents results of page walk overheads and TLB-reach problems for the die-stacked LLCs. Section VI concludes the study and lists the future work.

\section{x86-64 Page Walk and DRAM caches}
\subsection{Page Table Structure}
\begin{figure}
\centering
\subfigure[Page table structure in x86-64]{\includegraphics[scale=0.4]{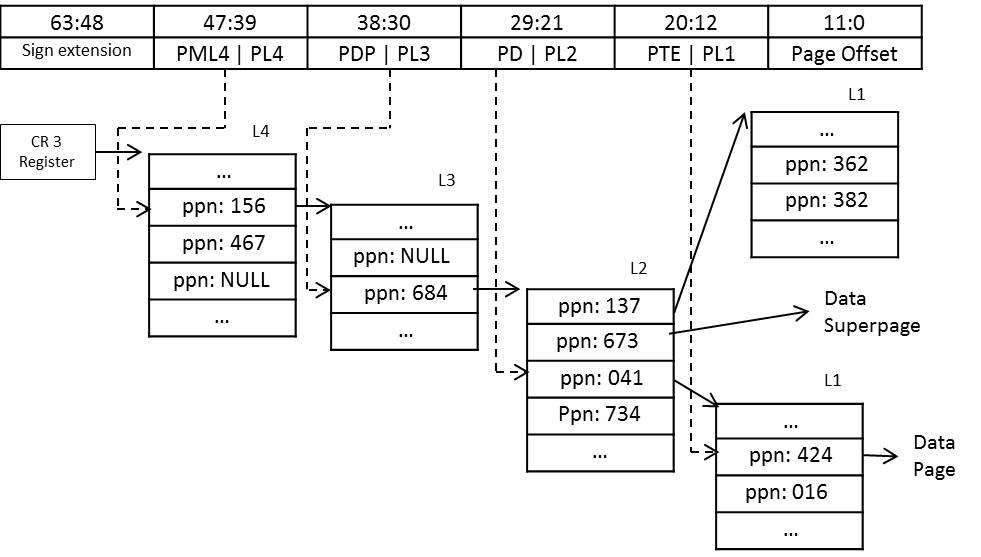}
\label{pagewalk}}
\subfigure[2D page walk in Virtualization]{\includegraphics[scale=0.4]{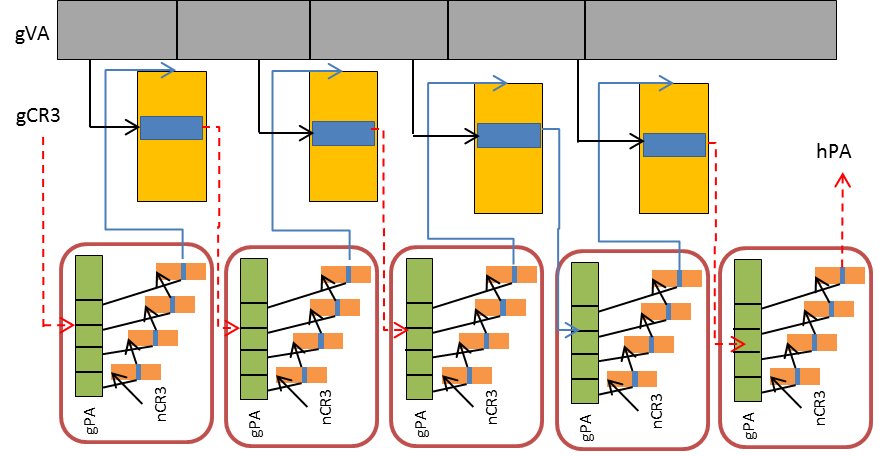}
\label{virtwalk}}
\caption{Page table structure in x86-64}
\label{pagetable}
\vspace{-1.5em}
\end{figure}

x86 processors use a radix tree to map virtual to physical address. The depth of the tree in modern processors currently stands at 4 levels to accommodate 48-bit virtual addresses. A virtual address is split into page offset and page number. The virtual page number is further divided into a sequence of indices. The first index selects an entry from the root level of the tree, pointed to by the CR3 register, which contains the pointer to the next level of the tree and so on until either the address is invalid (i.e. no valid translation is present for that address, in which case a page fault ensues) or the entry points to a data page. If the entry is found, the virtual address offset is concatenated with the physical page number to retrieve the complete physical address. In x86 systems the TLBs are \textit{hardware-managed}, meaning on a miss MMUs use hardware state machine to walk the page table and locate the mapping and insert to TLB. The effect of hardware-manged TLBs is that the page table organization is largely fixed but avoids expensive interrupts as in a \textit{software-managed} TLB \cite{softwareTLB}.\par

Figure \ref{pagewalk} shows the structure of a x86 64-bit virtual address. The standard page size 4KB is used, giving the 12-bit offset. The remaining bits are divided into 9-bit offsets to index into the four levels of the page table. The four levels are PML4(Page Map Level4), PDP(Page Directory Pointer), PD(Page Directory) and PT (Page Table) . For simplicity we will refer to these levels as PL4, PL3, PL2 and PL1 levels respectively. Given this structure of the page table, the current virtual address translation requires upto four memory references to "walk" the page table. Each entry at every level is 8 bytes (i.e. 512 PTEs fit in one 4KB page). Superpages, which are virtually and physically aligned pages of memory, were proposed to allocate large chunks of memory. Superpages of 2MB or 1GB can be allocated if the Operating System can find contiguous and aligned 2MB or 1GB regions in memory which are less likely in a long running system due to memory fragmentation. Superpages allow a single entry in the TLB to cover a larger memory address space, hence increasing the TLB-reach.\par

Figure \ref{arch} shows the architectural support for translation caching in modern processors which consists of multiple hierarchical levels of translation caching structures. Typically there are 64-entry, 4-way associative split L1 Data and Instruction TLB and a shared 8-way associative L2 TLB of 1024 entries for 4KB pages. Superpage TLBs contain a handful of entries for 2MB pages and 1GB pages which are looked up in parallel with L1 TLBs for a match. If these translation structures encounter a miss, accessing each page table level on every page walk will incur a penalty of several tens of cycles, even if all translations are present in the L2 data cache. To exploit the significant locality in the upper level entries, vendors have introduced low latency structures to store these upper level page table entries allowing the walk to skip a few levels. AMD's Page Walk Cache (PWC) and Intel's Paging-Structure Caches are examples of translation caches \cite{skipdontwalk}.

\subsection{Page Walk in Virtualization}
With virtualization, page walks could incur upto 24 memory references to complete a walk. This is due to the use of separate guest page tables (gPT) to translate guest virtual to guest physical addresses and nested page tables (nPT) to translate guest physical addresses to system physical addresses. Guest and nested page tables are set up and modified independently by the guest and hypervisor respectively and they have their own CR0, CR3, CR4, EFER and PAT state registers. When a guest attempts to reference memory using a virtual address and nested paging is enabled, the page walker performs a 2-dimensional walk using the gPT and nPT to translate the guest virtual address to system physical address as shown in Figure \ref{virtwalk}. When the page walk is completed, an entry containing the end to end translation is inserted into the TLB.

\subsection{Large Last Level Die-Stacked DRAM Cache}
The inclusion of large last level DRAM cache reduces off-chip memory access time. These DRAMs operate in the same model as off-chip DRAMs and are subject to the timing restrictions in terms of t$_{CL}$, t$_{RCD}$, t$_{RP}$ and t$_{RAS}$. The use of this capacity as cache at fine granularities incurs prohibitively high meta-data storage overhead. Prior research has worked towards reducing the tag storage overhead by co-locating tags and data (TAD) in the same row buffer and reducing cache hit latency by using a hit predictor \cite{alloycache}. Using these large caches could lead to a combination of TLBs hit/miss along with cache hit/miss. The inferences for the resulting combinations, in increasing order of latency, are listed in Table \ref{cachetlb}.

\begin{table}
\tiny
\centering
\caption{Cache access and TLB lookup}
\begin{tabular}{l|l|l|l}
\specialcell{Cache\\access} & \specialcell{TLB\\lookup}& Effect & Inference \\
\hline
HIT & HIT & data retrieved without additional delay & \specialcell{working set fits in cache \\ and TLB reach sufficient}\\
\hline
MISS & HIT & \specialcell{off-chip access for data,\\no overhead in translation} & \specialcell{TLB reach is sufficient,\\but exhibits poor cache locality} \\
\hline
HIT & MISS & \specialcell{multiple memory access for translation,\\but requested data resides in cache} & \specialcell{TLB reach insufficient for cache \\ translation dominates access time} \\
\hline
MISS & MISS & \specialcell{multiple memory access \\to retrieve both translation \& data} & \specialcell{possibly first access to the page / \\  no recent accesses / page fault}\\
\hline
\end{tabular}
\label{cachetlb}
\vspace{-1.5em}
\end{table}

\section{Related Work}
Many prior studies have pointed out the importance of TLB-reach and the necessity of speeding up miss handling methods. Prior work has proposed exploiting spatial locality by coalescing \cite{colt} and clustering \cite{clusteredtlb} a group of spatially contiguous TLB entries into a single entry in the TLB. On look-up, the offset between the base virtual address stored in the tag is used to calculate the offset from the base physical page. 
More recently there have been efforts to make the TLB structures more superpage friendly without unfairly biasing applications which use small pages. This has been achieved in \cite{superpagefriendly} using a combination of skewed TLB \cite{skewedtlb} and page size prediction on lookup. Abhishek Bhattacharjee et al. \cite{sharedtlb} propose last level shared TLB structures rather than the norm of per core private TLBs. \par
Direct segment \cite{directsegment} analyses the memory characteristics of big-memory workloads like database, memcached, graph500 etc. and observe that these workloads do not require fine grained protection or swapping as they are memory aware, long running and pay substantial cost for page walks. They propose primary region abstraction for these workloads which allows programs to specify a portion of their memory which does not benefit from paging and maps this region as a segment with base and offset register to completely eliminate paging overhead. Furthermore, in \cite{rmm} Jayneel Gandhi et. al propose allocation in OS in units of 'ranges' which are mapped using segment registers. Their scheme is backward compatible with traditional paging using a range TLB at L2 level which is looked up in parallel on a L1 TLB miss.\par
Virtualization has seen the widespread adoption in the cloud space, however, it comes with overheads in I/O and memory accesses due to expensive page walks using nested page tables. To reduce these overheads AMD \cite{amd2d}proposed accelerating 2D page walks using nested page walk caches which cache frequently used levels of the page walk in hardware, greatly reducing the overheads of TLB miss. Based on direct segment \cite{efficientvirt} seeks to futher reduce memory overheads of virtualization.\par
There have been efforts to characterize the behaviors and sensitivity of individual applications in the SPEC 2000 and PARSEC benchmarks in \cite{spectlb}  and \cite{parsectlb} respectively. For SPEC 2000 the effects of TLB associtivity, prefetching and super-paging was studied for each workload ignoring the OS involvement. For PARSEC workloads they explored the effect of sharing TLB structures with the idea that cores may fetch entries that maybe useful to other cores in the near future due to multi-threaded nature of the programs.\par
The other area of active research has been die-stacked DRAM organization. There have been significant efforts to reduce meta data overhead from various researchers world over. Most notably Loh-Hill cache \cite{lohhill} and direct mapped Alloy cache \cite{alloycache} organization use the TAD structure. More notably close to our work are TagTables \cite{tagtables} and Tagless DRAM \cite{taglessdram} cache that piggyback on the TLB translation mechanism to detect presence of the block in cache. Tagtables achieves this by flipping the page table organization, which as observed earlier is difficult to implement due to hardware managed TLBs. Tagless DRAM is more a practical solution which translates virtual address to a die-stacked cache address and moves the translation to physical address off critical path.

\section{Experimental Setup}

\subsection{Simulator Platform}
\begin{table}
	\caption{Experimental Setup}
	\begin{tabular}{@{}ll@{}}
		\toprule
		\textbf{Parameter}    & \textbf{Value}                               \\ \midrule
		Processor  & 3.9GHz, 4-core, 5way OoO, x86-64 ISA  \\
		L1 i-cache            & 32KB private, 4-way SA, 64B blocks, 2 cycles \\
		L2 d-cache            & 32KB private, 8-way SA, 64B blocks, 4 cycles \\
		L2 cache              & 256KB private, 8-way SA, 64B blocks, 6 cycles \\
		L3 cache              & 4MB shared, 12-way SA, 64B blocks, 9 cycles \\
		\midrule
		L1 iTLB 4KB pages     &  64 entries, fully associative   \\
		L1 dTLB 4KB pages     &  64 entries, fully associative   \\
		L2 TLB 4KB page		  & 1024 entries, unified fully associative \\
		Superpage TLB    	  & 32 entries, 2MB fully associative\\
		\tiny (as applicable) &\\
		\midrule
		Main Memory           & 4GB DDR2, 50ns                  \\
		\bottomrule
	\end{tabular}
	\label{setup}
	\vspace{-1.5em}
\end{table}
We use the QEMU based x86-64 cycle accurate full system simulator MARSSx86 \cite{marss86} for this study. We modify the simulators MMU to incorporate the various appropriately sized multi-level TLB structures. We add a shared L2 TLB, a dedicated superpage TLB for 2MB pages. We also modify the page walk handler to accurately model the reduced number of page walk levels for superpages. We boot unmodified Linux kernel 2.6.38 which supports Transparent Huge Pages (THP). We also include a L4 cache in the memory hierarchy and systematically account for timing characteristics in the simulator. This gives us the lower bound on the overheads. However, we leave the analysis of superpage TLBs as future work. Hence for this work, TLB insertions are done at 4KB page sizes. We chose full system simulation to be able to observe impact of TLB shootdowns due to context switches and influences of OS on TLB behavior, which has been shown to have considerable influence in the past literature \cite{softwareTLB}. \par
\begin{table}
	\centering
	\caption{Workloads}
	\begin{tabular}{ c ||c|c|c|c}
		\hline
		mix1 & milc & mcf & omnetpp & gcc\\
		mix2 & GemsFDTD & leslie3d & dealII & soplex\\
		mix3 & cactusADM & libquantum & tonto & shinpx3\\
		mix4 & lbm & bwaves & zuesmp & sjeng\\
		mix5 & milc & GemsFDTD & cactusADM & lbm\\
		mix6 & mcf & omnetpp & soplex & leslie3d\\
		mix7 & bwaves & astar & zeusmp & gcc\\
		mix8 & gobmk & bzip2 & h264ref & hmmer\\
		\hline
		\specialcell{PARSEC\\4-threads} &\multicolumn{4}{l}{canneal, dedup, ferret, fluidanimate, freqmine, raytrace}\\
		\hline
	\end{tabular}
	\label{workload}
	\vspace{-1.5em}
\end{table}

\subsection{Methodology}
In order to study the overheads of address translation we run a comprehensive set of multi-programmed and multi-threaded workloads. We selected memory-intensive benchmarks from the SPEC2006 suite \cite{spec2006} and those with large and unbounded working sets in PARSEC suite \cite{parsec}. The SPECCPU 2006 were compiled with 'base' tuning and run with 'ref' input data sets. Our workloads mix consists of four multi-programmed benchmarks, listed in Table \ref{workload}, prudently combined to create a representative mix of high, average and low memory activity. The workloads mixes were run for 4 Billion cycles in detailed simulation mode after fast forwarding the first 8 Billion instructions over the entire workload. In each case the total instructions executed for each configuration was more than 8 Billion instructions. In case of PARSEC applications we simulate the entire region of interest (ROI) of the application with 4 threads. The PARSEC applications were executed with the 'simlarge' input data set. Our baseline architecture parameters used for the simulation are listed in Table \ref{setup}, modeling the recent Intel Ivy Bridge processor. For die-stacked DRAM cache we simulate a flat cache with 16 way associative, 20 cycle latency writeback cache. L1 and L2 cache have 256 entry MSHR, while L3 and L4 have 128 entries each.

\section{Results}
\subsection{Page Walk latency analysis}
\begin{figure}
\centering
\subfigure{\includegraphics[scale=0.155]{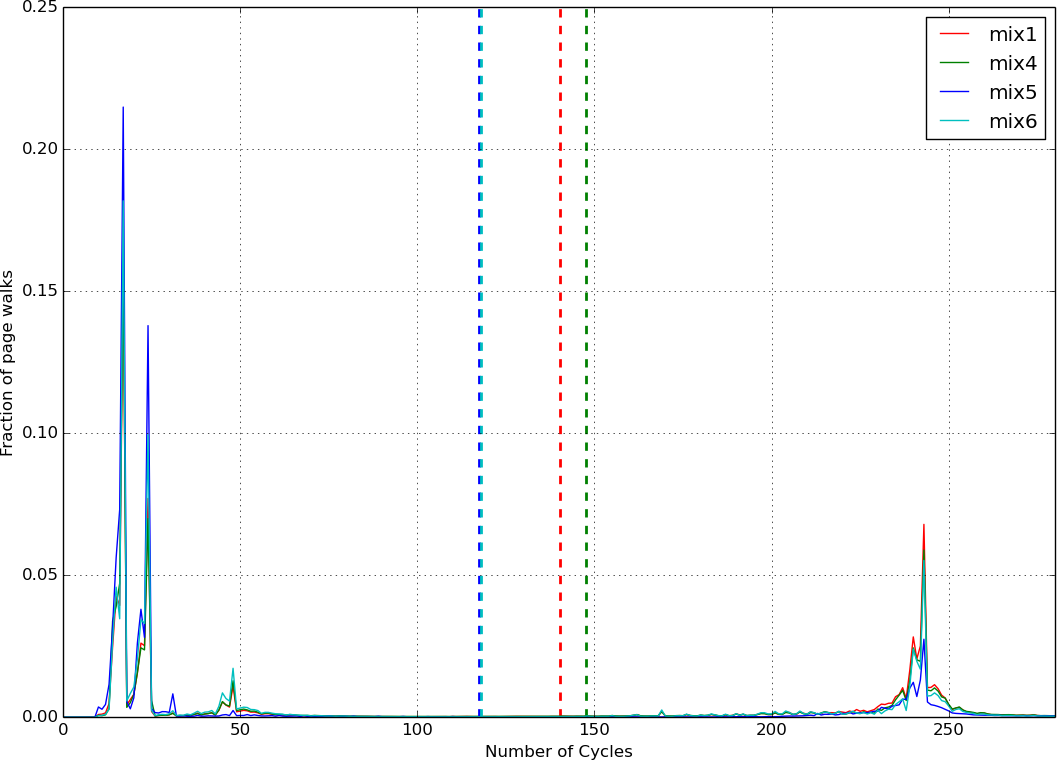}
\label{mix1456}}
\subfigure{\includegraphics[scale=0.155]{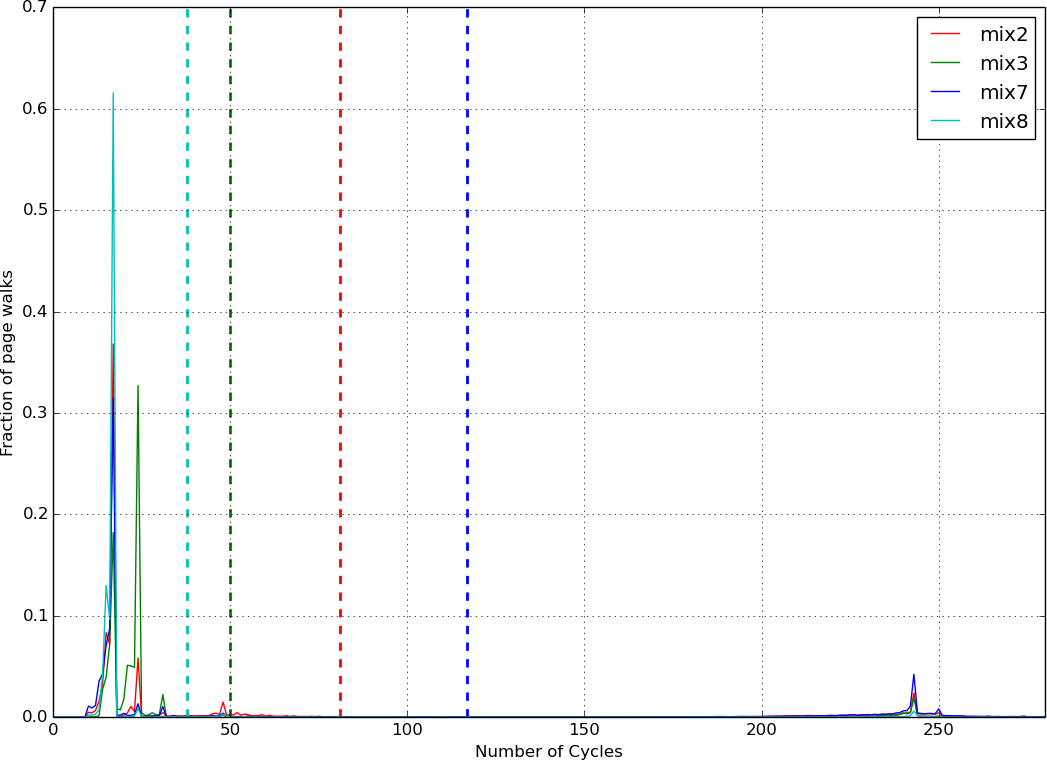}
\label{mix2378}}
\subfigure{\includegraphics[scale=0.25]{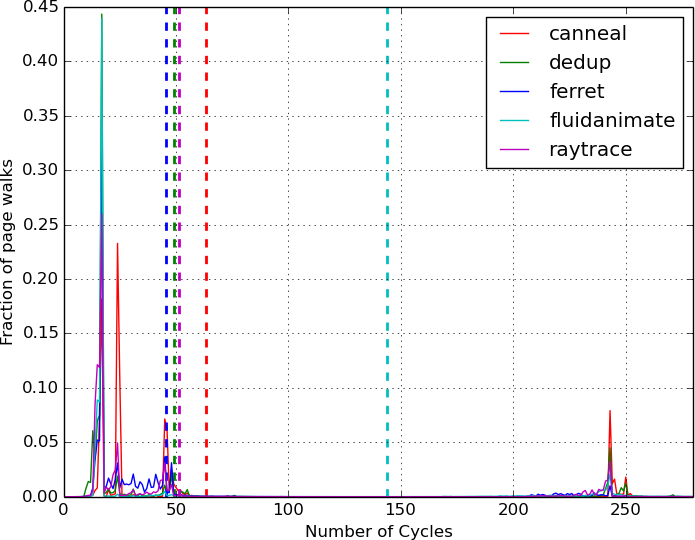}
\label{mix2378}}
\caption{Page Walk latency distribution for SPEC 2006 \& PARSEC workloads. The dashed bar shows the average latency}
\label{walklatencies}
\vspace{-1.5em}
\end{figure}
\begin{figure}
\centering
\renewcommand{\thesubfigure}{\tiny(\alph{subfigure})}
\subfigure[\tiny SPEC2006-mix1 to mix8 left to right]{\includegraphics[scale=0.215]{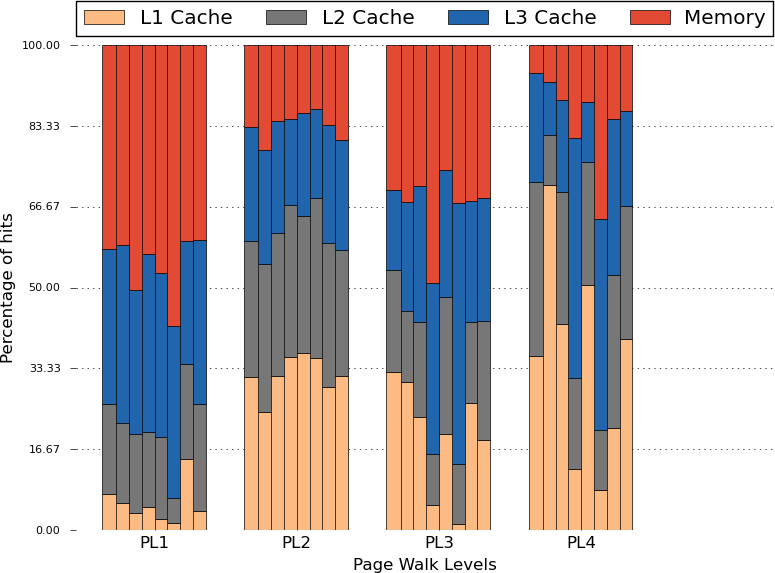}
\label{speccache}}
\subfigure[\tiny PARSEC-canneal, dedup, ferret, fluidanimate freqmine, raytrace - left to right ]{\includegraphics[scale=0.215]{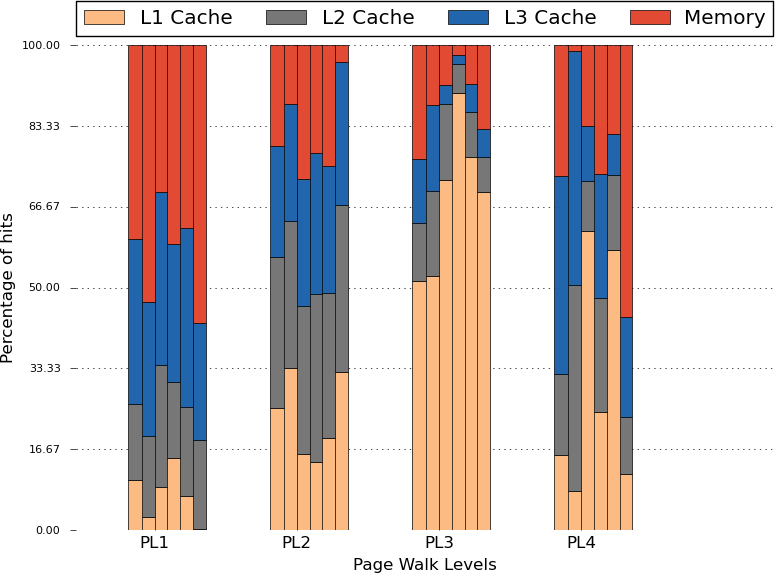}
\label{parseccache}}
\caption{Locality for each level of the page walk}
\label{walkcache}
\vspace{-1.5em}
\end{figure}

We limit our analysis to only data references (dTLB) in this work since data references are much less well-behaved than instructions in terms of misses. We find the iTLB hit rate to be greater than 99 \%  for most workloads. We studied the latency distribution incurred for a page walk to understand the average number of cycles taken to retrieve a translation on a TLB miss. Figure \ref{walklatencies} plots latencies experienced by page walk vs the fraction of walks that experience this latency. The plot shows that the workloads exhibit wide range of page walk latencies ranging from 20 cycles for mix8 and freqmine (not plotted) to 150 cycles for mix4 and fluidanimate. However, average page walk latencies are aggregate numbers for the entire page walk duration and are not adequate to conclude behavior of walks. TLB hit rates for most of the workloads were measured to be in the 90\% to 95\% range which makes page walks few and far spaced temporally. To better understand the page walk behavior we examined the locality characteristics of each page walk level.  To determine this we collected cache hit distribution of these page walk levels over the cache hierarchy of a multi-core architecture. Figure \ref{walkcache} shows the distribution of page walk hits for each workload. Clearly PL1 has highest memory access percentages amongst all workloads due to low locality. PL2 has a uniform hit percentage in almost all cache levels and we observe that most of the 8 translations in a cache line are used. For SPEC workloads PL3 level sees around 50 \% of the accesses are to either L3 or main memory. On the contrary for PL3, PARSEC sees very high L1 locality due to the fact that most PARSEC workloads have footprints of less than 1GB. Due to the cache pollution resulting from page walks AMDs MMU designs have made a design decision to walk page tables in L2 caches rather than L1 caches \cite{amd2d} which may exacerbate the walk latencies. \par
\begin{figure}
\centering
\captionsetup{justification=centering}
\renewcommand{\thesubfigure}{\tiny(\alph{subfigure})}
\subfigure{\includegraphics[scale=0.23]{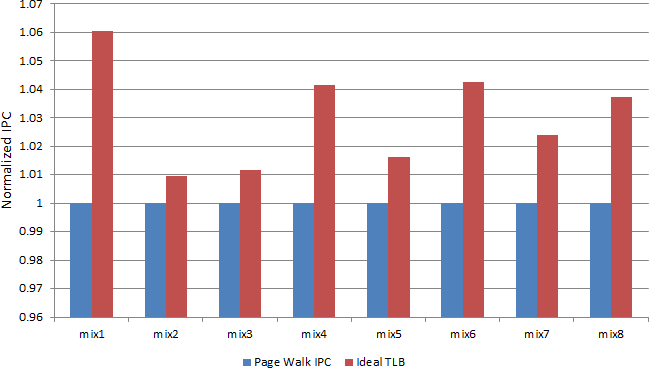}
\label{speccache}}
\subfigure{\includegraphics[scale=0.23]{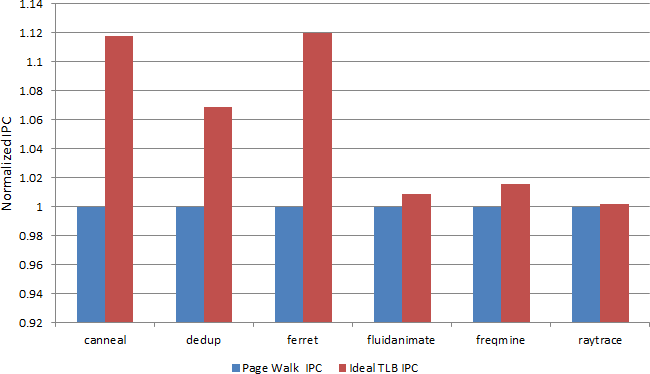}
\label{parseccache}}
\caption{IPCs normalized to baseline}
\label{walkipc}
\vspace{-1.5em}
\end{figure}

Modern OoO processors are very efficient at hiding latency of memory accesses with reorder buffers and load-store queues. Due to this masking, the latency incurred for page walks may not manifest as an overall IPC impact. To accurately understand the TLB miss impact on IPC we delve deeper and compare IPC against a simulated ideal TLB. We simulate an ideal TLB as (a) zero page walk overhead, (b) no cache pollution, (c) returns the translation in a single cycle. We compare the resulting IPC values to determine the precise page walk overheads. Figure \ref{walkipc} shows the IPC values normalized to the baseline performance. We observe that in some workloads, specifically mix1 and mix6 IPC increases by 6.05\% and 4.26\% respectively and for canneal and ferret the increase is 11.75 \% and 12.01\% compared to baseline. This shows that with larger memory footprint a reasonable IPC improvement is possible with improved address translation schemes.\par

\subsection{TLB reach in Die-Stacked DRAM Caches}
\begin{figure}[h!]
\centering
\includegraphics[scale=0.55]{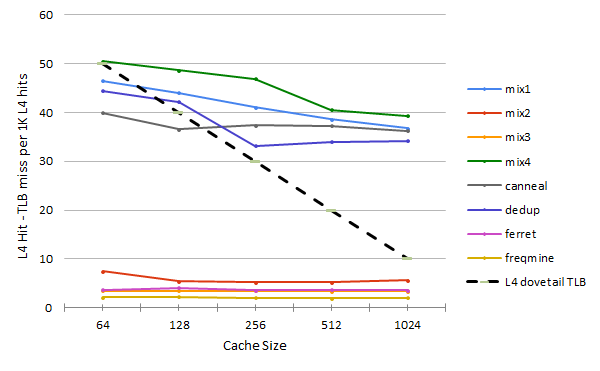}
\caption{L4Hit - TLBMiss per 1K L4 Hits - 64B line size}
\label{walkoverhead}
\vspace{-1.5em}
\end{figure}

To understand how often each of cases of cache and TLB hit-miss interplay occur in various workloads, we performed a detailed study for various cache capacities. We perform experiments for two different block sizes - 64B and 512B. Currently, we have been able to simulate 4 SPEC and 4 PARSEC workloads for the purposes of presentation in this paper. The rest of the configurations and workload permutations are currently being run and we expect to have results in the near future. Figure \ref{walkoverhead} plots the number of cases which result in L4 Hit-TLB miss per thousand L4 Hits for various cache sizes at 64B line size. We observe that as cache size increases this parameter almost flatlines out. For an L4 cache size tailored dynamic TLB reach, we should theoretically see a robust decrease in this parameter as shown by the \textit{L4 dovetail TLB} line. To complement the increase in L4 hit rate, a dynamic TLB would also need to correspondingly increase hits for the L4 cache to be effective. In other words for an oracle TLB this parameter would be 0.

\section{Future Work and Conclusion}
This is the first work that examines the TLB-reach and page walk overheads in the context of large last level die-stacked DRAM caches. We characterize the page walk latency and quantify effect on IPC. We layout the goals of an L4 dovetail TLB which adjusts reach as size of L4 cache increases to maximize efficiency of large die-stacked caches. Going forward we would like to examine the impact of superpage allocation, using dedicated superpage TLBs and vary TLB associativity. We also would like to simulate modern big data and cloud benchmarks with accurate latencies of DRAMs.





%

\end{document}